\documentclass[conference]{IEEEtran}  
\IEEEoverridecommandlockouts                              

\usepackage{amsmath} 

\usepackage{upgreek}
\usepackage[dvipsnames,table]{xcolor}

\usepackage{pgf}
\usepackage{subfigure}
\usepackage[export]{adjustbox}

\usepackage{algorithm}
\usepackage{algorithmicx,algpseudocode}
\usepackage[normalem]{ulem}
\graphicspath{{./figures/}}

\title{
Recurrent Neural Networks for Handover Management in Next-Generation Self-Organized Networks
}

\author{
\IEEEauthorblockN{
Zoraze Ali$^+$, Marco Miozzo$^+$, Lorenza Giupponi$^+$, Paolo Dini$^+$, Stojan Denic$^*$, Stavroula Vassaki$^*$
	}
\IEEEauthorblockA{
($^+$) Centre Tecnol\`ogic de Telecomunicacions de Catalunya (CTTC/CERCA), Barcelona, Spain \\ ($^*$) Huawei Technologies, Sweden AB \\ emails:\{zali, mmiozzo, lgiupponi, pdini\}@cttc.es, \{stojan.denic, stavroula.vassaki\}@huawei.com
}
}

\begin{document}

\maketitle
\thispagestyle{empty}
\pagestyle{empty}

\begin{abstract}
In this paper, we discuss a handover management scheme for Next Generation Self-Organized Networks. We propose to extract experience from full protocol stack data, to make smart handover decisions in a multi-cell scenario, where users move and are challenged by deep zones of an outage. Traditional handover schemes have the drawback of taking into account only the signal strength from the serving, and the target cell, before the handover. However, we believe that the expected Quality of Experience (QoE) resulting from the decision of target cell to handover to, should be the driving principle of the handover decision. In particular, we propose two models based on multi-layer many-to-one LSTM architecture, and a multi-layer LSTM AutoEncoder (AE) in conjunction with a MultiLayer Perceptron (MLP) neural network. We show that using experience extracted from data, we can improve the number of users finalizing the download by 18 \%, and we can reduce the time to download, with respect to a standard event-based handover benchmark scheme. Moreover, for the sake of generalization, we test the LSTM Autoencoder in a different scenario, where it maintains its performance improvements with a slight degradation, compared to the original scenario.
\end{abstract}



\date{Sept 2019}

\maketitle

\section{Introduction}
\label{sec:introduction}
It has been almost a decade since when Self-Organizing Networks (SON) have been defined and introduced as a feature of Long Term Evolution (LTE), in 3GPP Release 8. However, the vision of an automatic network capable of learning from experience and adapting to the environment has still not reached the maturity that operators were initially hoping, in order to maximize the efficiency of the network, while at the same time reducing the operational costs. 
5G cellular networks are characterized by extremely dense and heterogeneous deployments, in order to increase the network coverage and capacity. In addition, besides traditional sub-6 GHz and licensed bands, the access can span over a wide range of bandwidth, including mmWave and unlicensed spectrum. The high diversity of mobile devices and applications, further complicates the network architecture and its management. In this context, current and 5G networks generate a massive amount of measurements, control and management information~\cite{jessica:4Gto5G}\cite{Baldo:Bigdata}. This huge amount of information could be efficiently utilized to address the 5G network management challenges. Recently, the evolution in computational capabilities, has allowed to take advantage of machine learning and novel deep learning solutions to tackle multiple problems in different disciplines. In 5G and its evolution, the possibilities now available for machine learning and deep learning implementations are infinite and pave the way to an evolved vision of Next Generation SON to be able to address end-to-end solutions. 

The focus of this work is on the use case of handover management. In standards and literature, handover algorithms are traditionally based on standard events, e.g., the A3 or A2 event, and are mainly focused on the optimization of event trigger parameters, e.g., Hysteresis, Time-to-Trigger and Cell individual Offset \cite{Mwanje:Distributed}. Machine learning solutions have also been proposed in this respect, to adjust online these typical SON parameters~\cite{jessica:4Gto5G}. This approach presents the shortcoming that it considers the strongest signal for target cell selection before the handover, but not the actual perceived Quality of Experience (QoE) after the handover. For example, in urban scenarios where the handover to the strongest neighbour cell is successful but, a while after the handover, the transmission is deeply affected, e.g., by the presence of an obstacle, traditional handover approaches fail to provide a satisfactory solution, without taking advantage of available data to gain experience and make smarter decisions. Those approaches are likely to lead to a severe degradation of QoE, due to the unpredicted cell outage \cite{noms2016}.

In this paper, first we build a realistic cellular scenario using a high fidelity, full protocol stack, end-to-end network simulator, \textit{ns-3}, and we extract data from all the layers of the protocol stack. With this data, we build a wide and complete database, which we consider the basis of the experience that a smart network management should be able to construct. In order to make smart handover decisions in a scenario, in which we artificially generate deep zones of outage using obstacles, we use a Long-short-term-memory (LSTM) Recurrent Neural Network (RNN) to take advantage of the temporal characteristic of the data extracted from the network. The LSTM is designed in order to solve a regression problem to estimate the necessary time to download a file transmitted over a Transmission Control Protocol (TCP) transport. This is expected to capture the QoE of the users. We obtain very good prediction errors and with these results, we are able to prove that the learning approach outperforms traditional handover solutions. This means that once training is accomplished, the learning based handover algorithm is able to select a target cell for handover that could provide a better QoE, in the medium or long term, even if in the short term it offers a weaker signal upon handover decision. Moreover, we use an AutoEncoder (AE) with the purpose to compress the data and reduce its dimensionality. Then, this compressed data is used as input of an Feedforward Neural Network (FFNN), which offers excellent regression results similar to the one obtained using LSTM. This means, that the AE successfully reduces the dimensionality of the data without losing network performance. Finally, we show that the experience learned by these models in our scenario is also useful for making decisions in different deployment scenarios, so that the learned experience is proven useful to be reused in different geographical areas.

The outline of the paper is the following. In Section II, we discuss the target scenario and the data generation procedure. In Section III we introduce the handover control scheme. In Section IV we discuss the results of the learning scheme in comparison to traditional handover solutions. Finally, Section V concludes the paper.

\section{Data generation}
\label{sec:data generation}
\subsection{Simulation Scenario} \label{sec:simulation scenario}
\begin{figure}[!t]
\includegraphics [width=3.4in,height=2.5in]{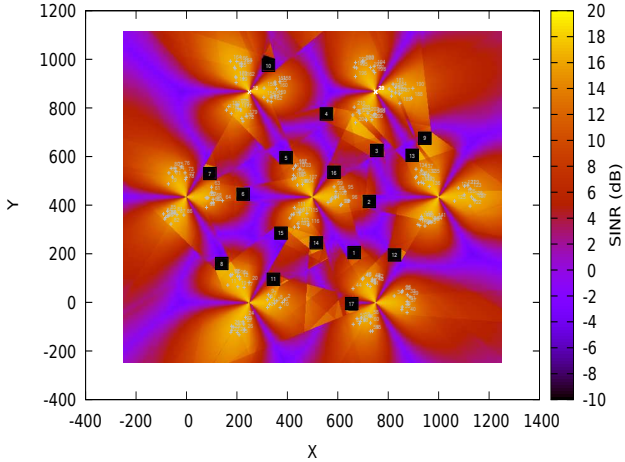}
\caption{Simulation scenario}
\label{fig_1}
\vspace{-0.4cm}
\end{figure}

\begin{table}[t]
\caption{Simulation parameters}
\label{table_sim_param}
\centering
\begin{tabular}{|>{\columncolor[gray]{0.95}}p{0.2\textwidth}|>{\columncolor[gray]{0.95}}p{0.2\textwidth}|}
\hline
\rowcolor[gray]{0.90}\textbf{Parameter} & \textbf{Value} \\
\hline
System bandwidth & 5 MHz\\
\hline
Inter-site distance & 500 m\\
\hline
Handover algorithm & A2-RSRP\\
\hline
Adaptive Modulation \&\, Coding Scheme & Vienna \cite{ns3}\\
\hline
SINR computation for DL CQI  & Control method \cite{ns3}\\
\hline
eNBs antenna type & Parabolic\\
\hline
eNBs antenna Beamwidth & 70 degrees\\
\hline
eNBs antenna max attenuation & 20 dB\\
\hline
Number of macro eNBs & 21 (7 cells)\\
\hline
eNBs Tx Power & 46 dBm\\
\hline
Distance between the center points of the cluster and the cell  & 100 m\\
\hline
Cluster diameter  & 50 m\\
\hline
Number of UEs in the system & 210 (30 per sector)\\
\hline
Mobility model & RandomWalk2dMobilityModel \newline Mode : Time \newline Speed : 10 m/s \newline Time : 40 sec \newline Distance : 4000 m\\
\hline
Path loss model & Cost231\\
\hline
eNB Antenna height & 30 m\\
\hline
Obstacle height & 35 m\\
\hline
Traffic & TCP Bulk File Transfer\\
\hline
File size & 1.5 MB\\
\hline
Maximum neighbours to handover  & 8\\
\hline
Total simulation runs & 20\\
\hline
Simulation time & 40 sec\\
\hline
\end{tabular}
\end{table}
\begin{table*}[]
\caption{List of measurements from LTE protocol stack used to create the dataset}
\label{table_input_features}
\begin{tabular}{|m{0.7cm}||m{5.3cm}|m{5.0cm}|m{5.0cm}|}
\hline
\multicolumn{4}{|c|}{\cellcolor[HTML]{C0C0C0}Input feature} \\ \hline
Layer & \multicolumn{3}{c|}{Measurements} \\

\hline
APP & \begin{tabular}[c]{@{}l@{}}1. Throughput UL\\ 5. Avg. number of rcvd packets DL\end{tabular} & \begin{tabular}[c]{@{}l@{}}2. Avg. number of rcvd packets UL\\ 4. Throughput DL\end{tabular} & \begin{tabular}[c]{@{}l@{}}3. Avg. number of rcvd bytes UL\\ 6. Avg. number of rcvd bytes DL\end{tabular} \\

\hline

RRC & \begin{tabular}[c]{@{}l@{}}7. Cell ID of serving cell\\ 10. Cell ID of neighbour 1\\ \hspace*{1cm} \textbf{.}\\ \hspace*{1cm} \textbf{.}\\ 31. Cell ID of neighbour 8\\ 34. Total number of radio link failures\end{tabular} & \begin{tabular}[c]{@{}l@{}}8. RSRP from serving cell\\ 11. RSRP from neighbour 1\\ \hspace*{1cm} \textbf{.}\\ \hspace*{1cm} \textbf{.}\\ 32. RSRP from neighbour 8\\ 35. Total number of handovers\end{tabular} & \begin{tabular}[c]{@{}l@{}}9. RSRQ from serving cell\\ 12. RSRQ from neighbour 1\\ \hspace*{1cm} \textbf{.}\\\hspace*{1cm} \textbf{.}\\ 33. RSRQ from neighbour 8\\ 36. First target cell ID to handover\end{tabular} \\
\hline

PDCP & \begin{tabular}[c]{@{}l@{}}37. Total number of txed PDCP PDUs DL\\ 40. Avg. PDCP PDU delay DL\\ 43. Min. PDCP PDU size DL\\ 46. Total number of rcvd PDCP PDUs UL\\ 49. Min. value of the PDCP PDU delay UL\\ 52. Max. PDCP PDU size UL\end{tabular} & \begin{tabular}[c]{@{}l@{}}38. Total number of rcvd PDCP PDUs DL\\ 41. Min. value of the PDCP PDU delay DL\\ 44. Max. PDCP PDU size DL\\ 47. Total bytes txed UL\\ 50. Max. value of the PDCP PDU delay UL\end{tabular} & \begin{tabular}[c]{@{}l@{}}39. Total bytes txed DL\\ 42. Max. value of the PDCP PDU delay DL\\ 45. Total number of txed PDCP PDUs UL\\ 48. Avg. PDCP PDU delay UL\\ 51. Min. PDCP PDU size UL\end{tabular} \\ 
\hline

RLC & \begin{tabular}[c]{@{}l@{}}53. Total number of txed RLC PDUs DL\\ 56. Total number of bytes received DL\\ 59. Max. value of the RLC PDU delay DL\\ 62. Total number of txed RLC PDUs UL\\ 65. Total bytes rcvd RLC PDUs UL\\ 68. Max. value of the RLC PDU delay UL\end{tabular} & \begin{tabular}[c]{@{}l@{}}54. Total number of rcvd RLC PDUs DL\\ 57. Avg. RLC PDU delay DL\\ 60. Min. RLC PDU size DL\\ 63. Total number of rcvd RLC PDUs UL\\ 66. Avg. RLC PDU delay UL\\ 69. Minimum RLC PDU size UL\end{tabular} & \begin{tabular}[c]{@{}l@{}}55. Total number of bytes txed DL\\ 58. Min. value of the RLC PDU delay DL\\ 61. Max. RLC PDU size DL\\ 64. Total bytes txed RLC PDUs UL\\ 67. Min. value of the RLC PDU delay UL\\ 70. Maximum RLC PDU size UL\end{tabular} \\
\hline

MAC & \begin{tabular}[c]{@{}l@{}}71. Initial MCS\\ 74. Avg. MCS UL\\ 77. Avg. RB occupied DL\\ 80. UL CQI\end{tabular} & \begin{tabular}[c]{@{}l@{}}72 Avg. TB size UL\\ 75. Avg. MCS DL\\ 78. DL CQI inband\end{tabular} & \begin{tabular}[c]{@{}l@{}}73. Avg. TB size DL\\ 76. Avg. RB occupied UL\\ 79. DL CQI wideband\end{tabular} \\
\hline

PHY & \begin{tabular}[c]{@{}l@{}}81. Avg. SINR DL\\ 84. Avg. number of UL HARQ NACKs\end{tabular} & 82. AVG. SINR UL & 83. Avg. number of DL HARQ NACKs \\
\hline

\multicolumn{4}{|c|}{\cellcolor[HTML]{C0C0C0}Output feature} \\ \hline
APP & \multicolumn{3}{l|}{1. File download time {[}sec{]}} \\ \hline
\end{tabular}
\end{table*}

We implement a realistic simulation scenario through ns-3 LENA LTE (Long Term Evolution) - EPC (Evolved Packet Core) simulator \cite{BaldoLena}. A macro cell outdoor scenario has been considered with a network consisting of three-sectorial eNBs. A cluster of UEs is placed in each sector at a fixed distance from the center of a cell, in which the UEs are dropped in random positions. Since, in this scenario, we use TCP as the transport protocol, such deployment of the UEs guarantees the establishment of a TCP connection between the remote host and the UEs. The UEs start moving after receiving the first packet, following a predefined mobility pattern. In order to increase the communication challenges in the scenario, and to generate more random coverage patterns, we introduce obstacles in the scenario, which generate multiple coverage holes, as shown in Fig.~\ref{fig_1}. Each UE is performing a TCP file transfer to a remote host in downlink and uplink direction. The complete set of simulation parameters are described in Table~\ref{table_sim_param}. The simulation consists of 20 runs of a deterministic handover to a potential neighbour. In this scenario, we observe that the maximum number of neighbours a UE manages to see is 8; therefore, each run is repeated 8 times to measure the QoE of a UE, i.e., file download time. For every simulation run, a UE picks a random starting position in the cluster and a random angle in the range of $[0^{\circ}$ to ~$360^{\circ}]$ to move away from the source eNB following a straight line. The data obtained from these deterministic handover campaigns for each UE is stored in the form of a dataset, according to the format described in the next subsection Sec.~\ref{sec:dataset creation}.

\subsection{Dataset creation} \label{sec:dataset creation}

We design our handover problem as a regression problem, where we need to estimate the QoE expected from performing handover to a particular target cell. In general, when working with supervised learning, such as in our case, one has to build a database to train, test, and evaluate the model. This dataset consists of input and output features stored in rows and columns. For this purpose, we extracted 84 measurements from each layer of the LTE protocol stack Table~\ref{table_input_features}. Hereafter, we mentioned them as input features. 3GPP already contemplates the upload of a part of these measurements, e.g., UE measurements, as part of the Minimization of Drive Test (MDT)~\cite{3gpp.36.331:ts} functionality. We gather measurements from all the layers of the protocol stack using some logs/traces already available in ns-3, e.g., those available for RLC (Radio Link Control) and PDCP (Packet Data Convergence Protocol), and other new custom trace sources at RRC (Radio Resource Control), MAC (Medium Access Control) and PHY, obtained by leveraging the tracing system of \textit{ns-3}. The input features, for our dataset, are extracted with the periodicity of 200 ms in order to be consistent with the approximate periodicity with which UE measurements are reported from UEs at the RRC level. This dataset can be expressed as a matrix $\mathbf{\overline {X}}$.
\begin{equation*}
\mathbf{\overline {X}} = 
\begin{bmatrix}
         \overline {x}_{1,1}&\overline {x}_{1,2}&\cdots &\overline {x}_{1,m} \\
         \overline {x}_{2,1}&\ddots&\cdots &\overline {x}_{2,m} \\
         \vdots & \vdots & \overline {x}_{i,j} & \vdots\\
         \overline {x}_{n,1}&\overline {x}_{n,2}&\cdots &\overline {x}_{n,m}
\end{bmatrix}
\end{equation*}
where the feature vector of size 84 is $\overline {x}_{i,j} \in \mathbf{\overline {X}}$, $ 1 \leq i \leq n$, \ and $1 \leq j \leq m$, with $1 \leq n \leq 33600$ , $m = 200$.

The total number of samples, i.e., the upper limit of $n$, can be computed by multiplying the total number of UEs with the maximum neighbours to handover, and the total number of simulation runs (see Table~\ref{table_sim_param}). On the other hand, $m$ corresponds to the number of time-steps that the LSTM processes to infer the time to download by a UE, which is 200 in our case.

\section{LSTM models for handover management}
\label{sec:handover control}
As mentioned in Section \ref{sec:data generation}, the dataset consists of the measurements and traces extracted periodically from each layer of LTE protocol stack, forming a time series of multivariate features. We believe that, by exploiting the temporal characteristic of this data one could understand the impact of handover decisions. Therefore, in this paper we employ RNN with LSTM units \cite{lstm1}. It is a special kind of RNN, which outperforms other machine learning approaches for time series analysis \cite{Wang:BigDataMl}~\cite{trinh:lstm}, and solves the problem of long-term dependency issue found in vanilla RNN~\cite{lstm2}.

\begin{figure}[!b]
\includegraphics [width=3.4in,height=2in]{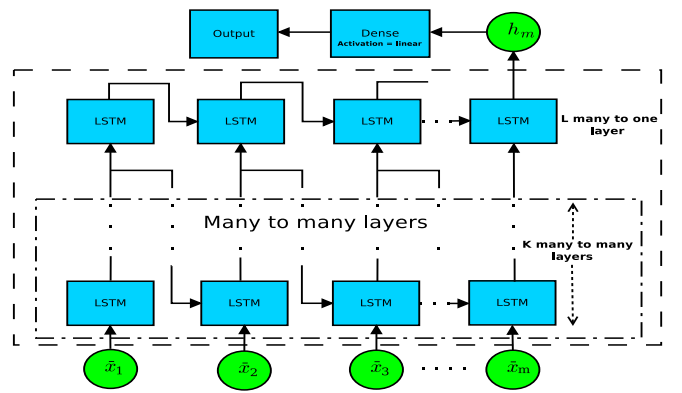}
\caption{Proposed model 1 : Many to one LSTM architecture}
\label{fig_2}
\end{figure}

\begin{figure}[!b]
\includegraphics [width=3.4in,height=2.3in]{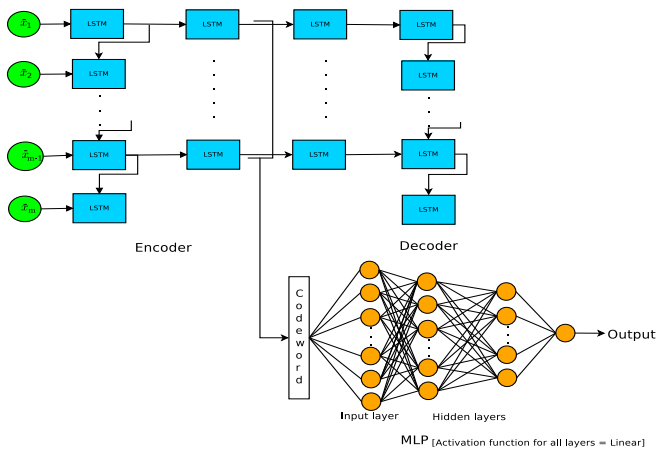}
\caption{Proposed model 2 : Autoencoder + MLP neural network}
\label{fig_3}
\end{figure}

In this paper, we present two models to predict the time required by a UE to download a file using the dataset. Fig.\ref{fig_2}, shows the first model, which is based on a multi-layer many-to-one LSTM architecture. This model takes 200 timesteps (i.e., $m$), each comprises of 84 features, as input, and process them in a single lag with multiple batches of size 32, to infer the time to download. On the other hand, the second model is based on a multi-layer LSTM AE \cite{NIPS2015_5949} in conjunction with a MultiLayer Perceptron (MLP) neural network, as shown in Fig.\ref{fig_3} . An AE is an unsupervised machine learning algorithm, which learns a function to approximate an output identical to the input. Since it is based on the \textit{encoder-decoder} paradigm, the input is transformed into a lower-dimensional space, also known as Codeword (CW), to more efficiently model highly non-linear dependencies in the inputs. The compression operation manages to extract more general and useful features, which retain important aspects of a dataset~\cite{Masci:Stacked}. Our goal is to smartly reduce the data to be used for inferring the time to download.

\begin{figure}[!t]
\includegraphics [width=3.4in,height=2.3in]{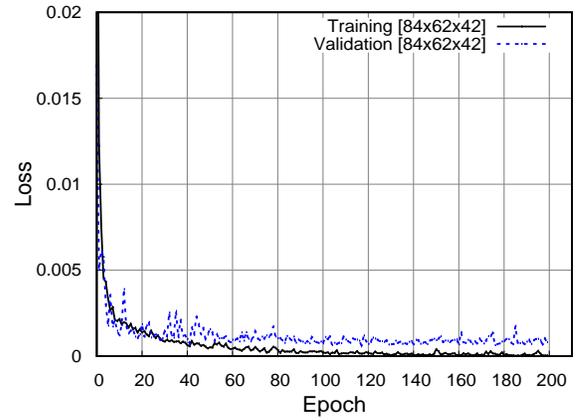}
\caption{Mean square error per epoch of LSTM [84x62x42]}
\label{fig_4}
\end{figure}

\begin{figure}[!b]
\includegraphics [width=3.4in,height=2.3in]{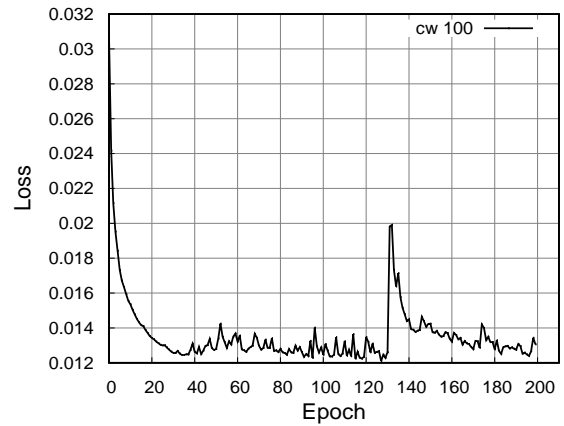}
\caption{Mean square error per epoch of AE (CW = 100)}
\label{fig_5}
\end{figure}

\begin{figure}[!t]
\includegraphics [width=3.4in,height=2.3in]{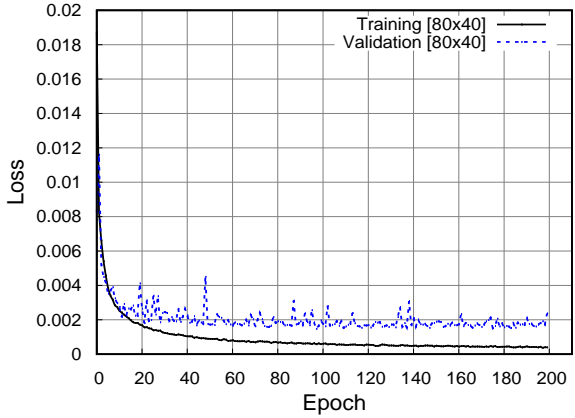}
\caption{Mean square error per epoch of MLP [80x40]}
\label{fig_6}
\end{figure}

\section{Performance Evaluation}
\label{sec:perf eval}
The implementation of the proposed models is done in Python, using Keras and Tensorflow, as backend. In particular, to speedup the training, testing, and evaluation of these models we used fast LSTM implementation with Nvidia CUDA Deep Neural Network (CuDNN) library for GPUs~\cite{chollet2015keras}. We note that, to select the hyperparameters of the first model, i.e, number of layers and the number of LSTM units (blue LSTM blocks in Fig.~\ref{fig_2}) in each layer, we tested nine different combinations. Finally, the hyperparameters resulted in a lowest average Mean Square Error (MSE) (over 200 epochs) were selected. For the second model, we use a similar approach first to select the CW length of the AE, among five different CW lengths. Then, using this selected CW as an input to the MLP neural network, one set of hyperparameters, among 7, was chosen for the MLP. For the readability purpose, Fig.~\ref{fig_4}, Fig.~\ref{fig_5}, and Fig.~\ref{fig_6} show the loss, i.e., Mean Square Error (MSE) per epoch, only for the selected hyperparameters. In particular, Fig.~\ref{fig_4} shows the MSE per epoch of the first model using 3 (i.e., \textbf{L = 1, K = 2}) layers of LSTM nodes, where the numbers separated by ``x" represent the number of hidden LSTM units. We observe that, after 140 epochs, this model is able to achieve and maintain very low testing loss independently from the number of layers and cells per layer.  Related to the second model, Fig. \ref{fig_5} shows the AE loss using CW length of 100 over 200 epochs. We note that, this loss is the MSE between the decoded version of the input data and the original input data fed to the AE. Similarly, Fig. \ref{fig_6} shows training and validation loss of the final, two layered (80 and 40 neurons in first and second hidden layer, respectively) MLP neural network fed with the input of CW length of 100. 

\begin{figure}[!t]
\includegraphics [width=3.4in,height=2.5in]{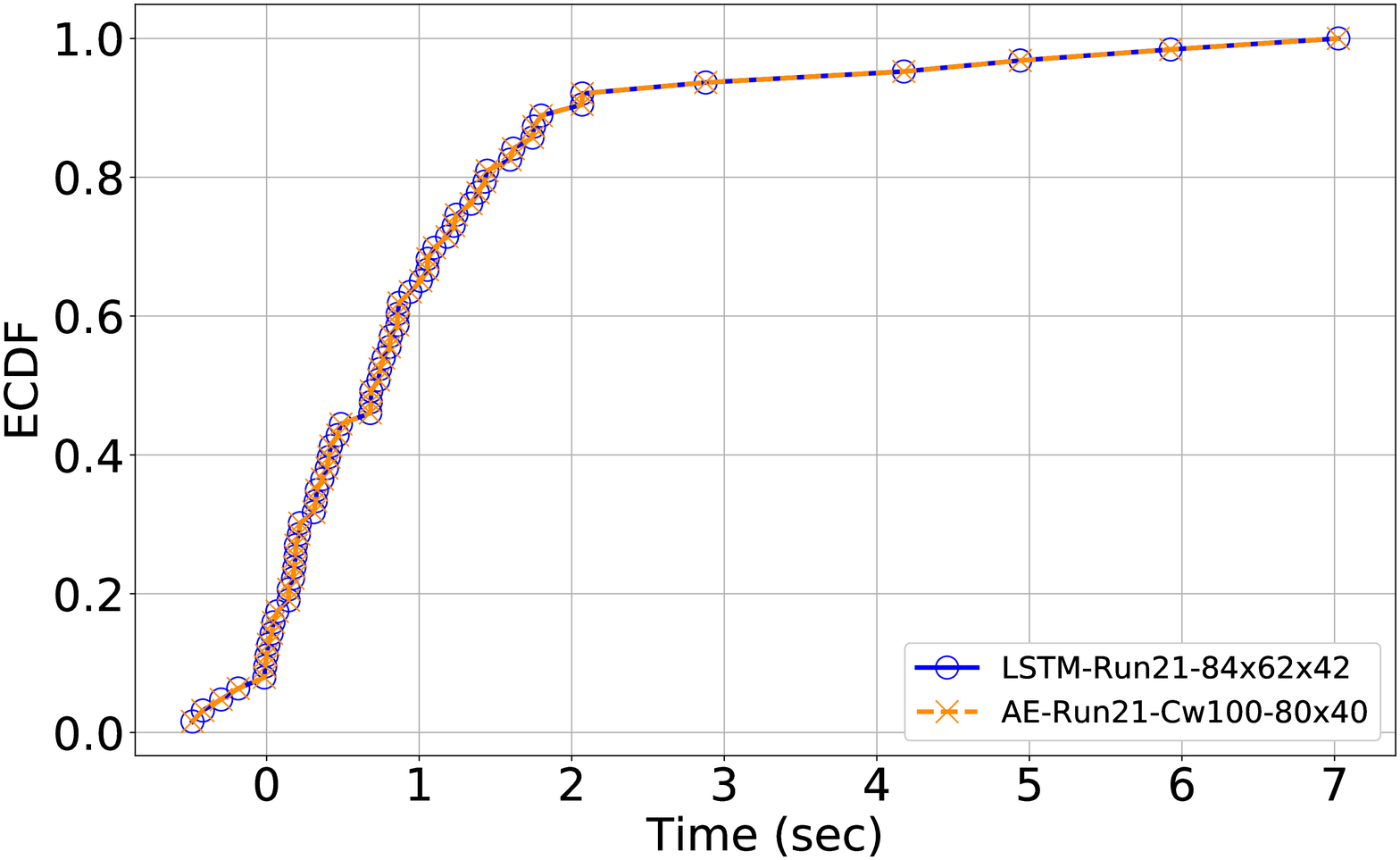}
\caption{The difference between SOTA and ML DL time for common UEs}
\label{fig_7}
\end{figure}

The performance evaluation of these models is performed in an offline fashion, i.e., by comparing the real time to download for each UE, obtained after selecting the target cell providing the lowest predicted time to download, to the one achieved by using a benchmark approach, i.e., A2-RSRP based handover algorithm. In particular, to perform this evaluation we consider another dataset generated with two extra simulation campaigns using a Run value which was not used to build the training dataset (i.e. Run 21). The first campaign aims at gathering the file download time using the benchmark handover algorithm (e.g., A2-RSRP).The second simulation campaign is conducted in a similar way as the one to build the training dataset, i.e., it consists of 8 deterministic handovers. Following this approach, we construct 8 input strings, for each neighbour of a UE, which consists of 1 row and 16800 columns (i.e, 84 features x 200 time steps). To obtain a predicted time to download for the selected LSTM and AE architectures, these strings are used individually as their input. Finally, for each UE, we select the eNB with the minimum predicted time to download for the handover. The results obtained using the benchmark handover algorithm show that there are 63 UEs out of 210, which are able to finalize the download. On the other hand, using the two trained models, 77 UEs are able to download the file successfully. This means that the machine learning approach manages to increase by 18 \% the number of UEs, which are able to finalize the download during the simulation time. Moreover, there are 62 common UEs, which were always able to download the file, irrespective of the tested handover solution, i.e., benchmark, LSTM or AE based. Fig.\ref{fig_7}, plots the Empirical Cumulative Distribution Function (ECDF) of the difference between the download time observed by these UEs using the benchmark, and the two proposed models. The ECDF trend, on the positive x-axis shows that, using LSTM or AE we are able to reduce the file download time for 56 UEs compared to the benchmark case. However, there are 6 UEs which experience marginally higher download time compared to the benchmark (see the trend on -ve x axis in Fig.\ref{fig_7}).
\begin{figure}[b]
\includegraphics [width=3.4in,height=2.5in]{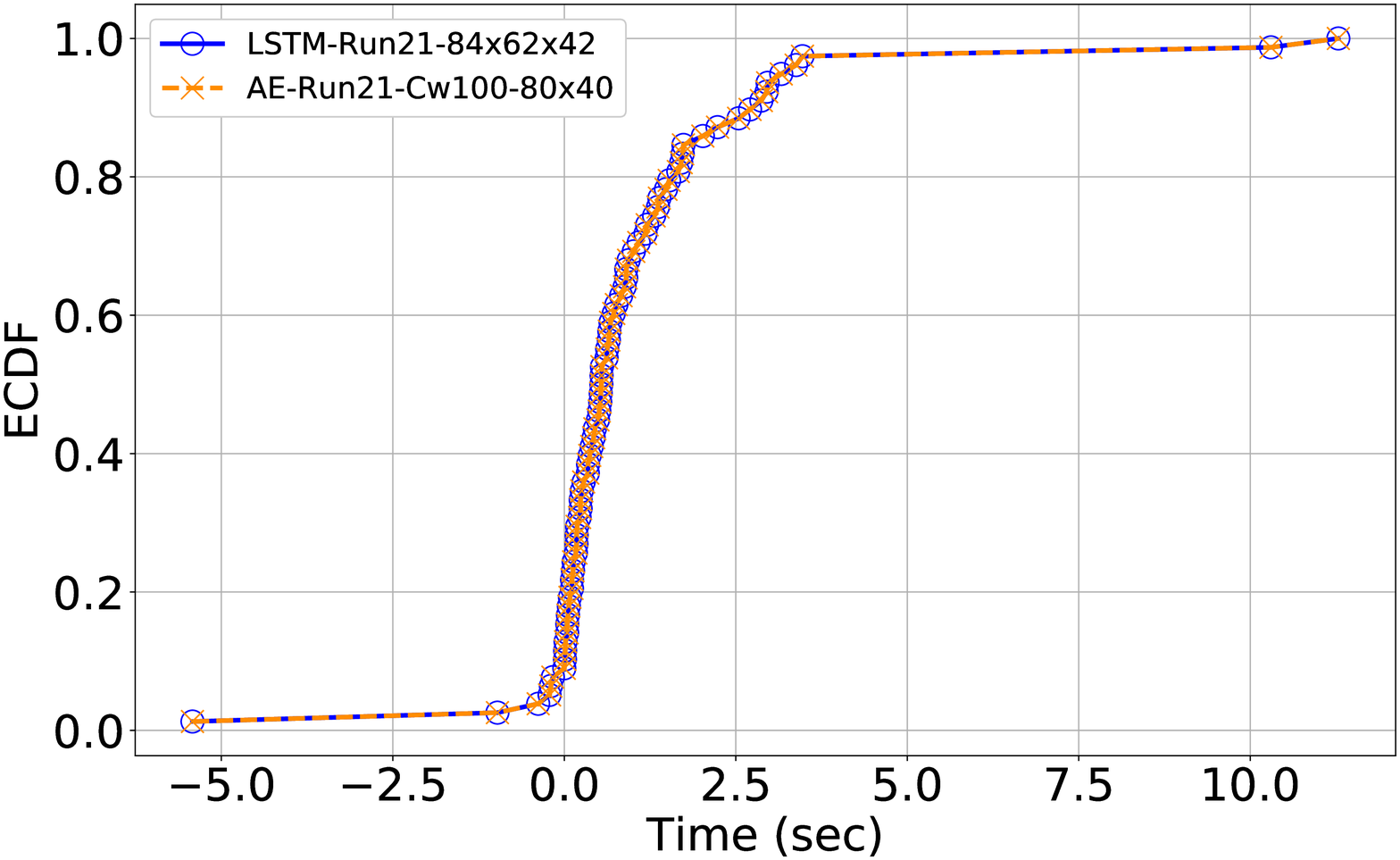}
\caption{The difference between SOTA and ML DL time for common UEs}
\label{fig_8}
\end{figure}
We believe that their performance can be improved by increasing the size of the database used to train the models and by further fine tuning their hyperparameters. Moreover, this evaluation shows that the MLP, fed with the AE CW of 100 performs similarly to the LSTM. This proves that the AE has efficiently transformed the inputs into a lower-dimensional space without losing the meaningful information of the dataset for the use case of the handover. To further investigate the reusability of these trained models, we test them in a simulation scenario with different deployment of the obstacles. In this scenario, using the benchmark handover algorithm 78 UEs out of 210 are able to download the file. On the contrary, the two models perform similar to each other, and offer an increase of 11.3636 \%, (i.e., 88) in the number of UEs, which are able to finish the download. Similarly, the ECDF in Fig.\ref{fig_8} shows that using the two models, out of 77 common UEs, we are able to decrease the file download time for 71. However, due to the presence of new temporal data introduced by the new spatial characteristic of the outages, there are 6 UEs, which experience high download time. This can be recovered by extending the available knowledge obtained in the old scenario, with new incremental data from the new scenario.

\section{Conclusions}
\label{sec:conclusions}
In this paper, we have exploited heterogeneous data, which is already available inside the network at different layers of the LTE protocol stack. This data is used to gain meaningful experience to make handover decisions, which are not based on the signal strength before the handover, but on the expected QoE after the handover. We have first proposed an RNN that exploits the temporal characteristic of this data. In particular, an LSTM is designed to model a regression problem that estimates the expected time to download a file for the different neighbours of the current serving cell. Our approach outperforms a traditional event-based benchmark handover scheme in terms of the number of successful downloads and time to download statistics. To reduce the dimensionality of the data and then facilitate the possibility to transfer the experience inside the network, we have proposed also a second model based on an LSTM-AE to compress the data up to a codeword of 100 and then an MLP neural network that implements the regressor. We tested this model in two different simulation scenarios, and conclude that its performance is equivalent to the one achieved using the LSTM trained with uncompressed data. Moreover, this also encourages us to go one step further to extend our work in the future, where we could leverage AE based multitask learning using the same database.

\section*{Acknowledgements}
This work was partially funded by Spanish MINECO grant
TEC2017-88373-R (5G-REFINE) and Generalitat de Catalunya grant 2017 SGR 1195. It was also supported by Huawei Technologies, Sweden AB. 
\bibliographystyle{IEEEtran}
\bibliography{IEEEabrv,Recurrent_Neural_Networks_for_Handover_Management_in_Next-Generation_Self-Organized_Networks}

\end{document}